# Observation of Light Guiding by Artificial Gauge Fields

Yaakov Lumer[1†], Miguel A. Bandres[1†], Matthias Heinrich[2], Lukas Maczewsky[2],

Hanan Herzig-Sheinfux[1], Alexander Szameit[2], Mordechai Segev[1]

[1] Physics Department and the Solid State Institute, Technion – Israel Institute of Technology, Haifa 32000, Israel

[2] Institut für Physik, Universität Rostock, D-18051 Rostock, Germany

†These authors contributed equally to this work.

**The use of artificial gauge fields enables systems of uncharged particles to behave as if affected by external fields. Generated by geometry or external modulation, artificial gauge fields have been instrumental in demonstrating topological phenomena in many physical systems, including photonics, cold atoms and acoustic waves. Here, we demonstrate experimentally for the first time waveguiding by means of artificial gauge fields. To this end, we construct artificial gauge fields in a photonic waveguide array, by using waveguides with nontrivial trajectories. First, we show that tilting the waveguide arrays gives rise to gauge fields that are different in the core and the cladding, shifting their respective dispersion curves, and in turn confining the light to the core. In a more advanced setting, we demonstrate waveguiding in a medium with the same artificial gauge field and the same dispersion everywhere, but with a phase-shift in the gauge as the only difference between the core and the cladding. The phase-shifted sinusoidal trajectories of the waveguides give rise to waveguiding via bound states in the continuum. Creating waveguiding and bound states in the continuum by means of artificial gauge fields is relevant to a wide range of physical systems, ranging from photonics and microwaves to cold atoms and acoustics.**

Waveguiding – the ability to confine light in a region and guide it within a structure is a fundamental building block in many photonics applications. The basic structure of a waveguide consists of a core region where the light is confined and a cladding region within which the core is embedded. Standard waveguide structures rely on total internal reflection, and are made up of a high refractive index core surrounded by a low refractive index cladding, or on metallic pipes where the electromagnetic fields are completely forbidden from escaping [1]. Other schemes rely on photonic systems with bandgaps [2–6], coupled resonator arrays [7], grating-mediated waveguiding [8], Kapitza-effect arrangements [9] or exploit the vectorial spin-orbit interaction of light in an anisotropic medium [10]. In all of these, the core and the cladding are made from media with a different dispersion relation. Recently, a conceptually new waveguiding mechanics was proposed: a waveguide where the cladding and core region have the same dispersion curve, but the core's dispersion is shifted by virtue of an artificial gauge field [11].

Gauge fields are a fundamental concept in physics, describing the basic interactions between charged particles. Neutral particles (such as photons) are thus decoupled from real gauge fields. However, by properly engineering a physical system, one can generate artificial gauge fields that would govern the effective dynamics of the neutral particles. That is, most often an artificial gauge field can be induced through the geometric design of the system or through some specific external modulation, such that the effective dynamics of the system behaves as if it were governed by a real gauge field. Nowadays, artificial gauge fields play a crucial role in physics, since they allow us to endow systems with a wide range of intriguing features that are "naturally" not expected in them. From the fundamental side, this opens the door to explore new physics, while from the applications point of view, it facilitates new devices by enabling control over the dynamics of systems in new ways. For example, in cold neutral atoms it has been demonstrated that a synthetic magnetic field can be induced by rotating the system [12] or by judicious optical coupling between internal states of the atoms [13]. Furthermore, artificial gauge fields made it possible to explore topological phenomena outside the context of electronic systems, in photonics [14,15], ultra-cold atoms [16,17], opto-mechanical systems [18], and even in acoustic [19,20] and mechanical systems [21,22]. This is important since bringing topological protection to virtually all wave systems opens the door to a wealth of applications never envisioned before.

In photonic systems specifically, artificial gauge fields were proposed and demonstrated to induce dynamics that would otherwise be inconceivable for light. For example, artificial gauge fields were proposed in coupled resonators schemes as a means to realize wave dynamics under an effective magnetic field [23–26]. Numerous experiments followed, ranging from the induction of a strong effective magnetic field by spatially ordering waveguide arrays [27] and the creation of synthetic electric fields via making the waveguides helical – as was used to create photonic Floquet topological insulators [14] – to specifically-designed bianisotropic photonic crystals that give rise to topologically robust propagation [28]. In similar vein, dynamic localization [29–33] and Aharonov-Bohm phases were demonstrated [34], as well as non-reciprocal devices using temporally modulated silicon photonics [35]. In this spirit of exploiting artificial gauge fields, novel optical devices have been proposed such one-way mirrors, negative refraction, and highly efficient mode convertors [36,37]. Finally, the concept of artificial gauge fields was introduced to active (non-Hermitian) systems, as highlighted by the recent observation of topological insulator lasers [38,39].

Recently, Lin & Fan proposed to employ artificial gauge fields to induce a fundamentally new kind of waveguiding mechanism [11]. The proposed structure consists of a core and cladding regions that have the same underlying dispersion relation, but are subject to different artificial gauge fields, which in turn shift their dispersion relation with respect to one another. This shift between the dispersion curves in different regions makes it possible to design a structure where the core supports guided modes that do not have matching propagating states in the cladding region. Thus, the wavefunctions of these modes decay exponentially in the cladding regions, while being oscillatory in the core – as guided modes always are. To implement this idea, it was proposed to use arrays of resonators with a refractive index temporal modulations on different resonator. However, thus far the concept of guiding light by artificial gauge has never been demonstrated in experiments.

Here, we present the first experimental realization of waveguiding by artificial gauge fields. We implement this idea in an array of evanescently coupled identical waveguides, where the waveguides in the core and cladding follows different trajectories during propagation. In our system the artificial gauge field stems solely from the trajectory of the waveguides, and does not require external modulation of the material properties. Specifically, we present theoretically and experimentally two mechanisms for waveguiding by artificial gauge fields. In our first realization, the dispersion relation in the core and in the cladding are shifted from one another in momentum-space because of the artificial gauge field. This engineered dispersion relation creates propagating modes that are confined to the core only in specific ranges of transverse momenta. Our second realization is even more intriguing: the system is made not only of the same material, but also has the same artificial gauge field, such that the core and cladding have exactly the same dispersion relation. In that system, we induce guiding strictly by introducing a phase-shift between the gauge field of the core and the gauge field of the cladding. We find that this system exhibits leaky modes for all values of transverse momentum, except for a specific value where perfect core confinement is obtained. This specific value can be controlled by the system parameters – i.e. the waveguide trajectories. We show that this effect constitutes *a dynamic bound state in the continuum (BIC)* [40–44]. Thus, our experiments are in fact the first observation of dynamic BIC ever. Altogether, the ideas demonstrated here in our photonic platform can be implemented in other physical systems – such as ultracold atom systems, mechanical and acoustic systems – thereby enabling novel types of physical configurations, such as waveguides, couplers and photonic networks in previously unconceived scenarios.

We begin by describing our system: a photonic lattice of evanescently coupled waveguides. The propagation of light in this structure is described by the paraxial wave equation

$$i\partial_z \psi(\vec{r}) = -\frac{1}{2k_0}\nabla_\perp^2 \psi(\vec{r}) - \frac{k_0}{n_0}\Delta n(\vec{r})\psi(\vec{r}), \quad (1)$$

where $\psi$ is the envelope of the electric field, $k_0$ is the wave number in the bulk material, $n_0$ is the ambient refractive index, $\nabla_\perp^2 = \partial_x^2 + \partial_y^2$ and $\Delta n(\vec{r}) = n(\vec{r}) - n_0$ is the relative refractive index profile. Our basic building block is a one dimensional array of evanescently coupled waveguides, which means that $\Delta n(\vec{r})$ is a periodic function in $x$ with period $d_x$ such that each period consists of a single waveguide. Using a coupled mode theory approach [45,46], we can show that the spectrum of such an array is $\beta = 2c_1 \cos(k_x d_x)$ where $k_x$ is the spatial momentum in the $x$ direction, and $c_1$ is the coupling strength between neighboring waveguides (taken to be a positive number). Here, we define the propagation constant $\beta$ by the ansatz $\psi(x,y,z) = \psi_0(x,y)e^{-i\beta z}$. Titling the array at an angle $\eta$, makes the waveguides follow the trajectory $x - \eta z = const$. The spectrum of such a tilted array is [46]

$$\beta(k_x) = 2c_1 \cos\big((k_x - k_0\eta)d_x\big) + \eta k_x. \quad (2)$$

We notice that the tilting of the waveguides has two effects. First, the shift of $k_0\eta$ in $k_x$ results from the constant tilt of the waveguides, and constitutes a type of Gallilean transformation. Second, the shift of $\eta k_x$ in the spectrum results from the fact that there is an angle between the $z$ axis and the waveguides. Looking at Eq. (2), we can treat the tilt of the array as the effect of an artificial gauge field that depends on the tilting angle $\eta$. When we generalize and consider a two-dimensional array of waveguides, all tilted with the same angle $\eta$, we can treat our system as being subjected to a uniform gauge field. Such a uniform tilt would constitute a trivial gauge, that is, a trivial change of reference frame for the entire system which can be readily gauged away. To generate a non-trivial gauge-field, we couple arrays that have different tilting angles. In such a system, the gauge field becomes space-dependent and nontrivial, i.e., there is no reference frame in which both arrays are not tilted. More importantly, the band structures of two arrays of different tilt angles are not the same due to the different gauge field, even though the arrays are identical up to a tilt. In this way, by modifying the band structures of the arrays by means of artificial gauge fields, we can now design a structure that exhibits guiding of light by virtue of an artificial gauge field, in the spirit of Lin & Fan [11].

Consider a two-dimensional array of evanescently coupled waveguides, with the middle rows acting as the core and the rest are acting as the cladding as depicted in **Fig. 1a,b**. The rows of waveguides in the core and the cladding are identical in every parameter (refractive index, waveguide shape, distance between the waveguides), except the tilting angle – the core is tilted with at an angle of $+\eta$ and the cladding is tilted at an angle of $-\eta$, as shown in **Fig. 1b**. Thus, the core and the cladding regions are subjected to different artificial gauge fields. Notice that the gauge field is nontrivial, as the gauge field gradient specifically at the interface region cannot be gauge away by a change of reference frame. Thus, although our system is still periodic in $x$ with period of $d_x$, it is no longer $z$–independent; rather, it is periodic in $z$ with a period of $Z = d_x/\eta$. We calculate the spectrum of this photonic lattice (**Fig. 1a**) by Floquet diagonalization of the continuous paraxial equation Eq. (1) [46], this can also be calculated using an approximate tight-binding model [46]. The spectrum is presented in **Fig. 1c**. One immediately notes that the spectrum is not symmetric with respect to $k_x$ – a fact which directly stems from the $z \rightarrow -z$ parity symmetry breaking of the underlying structure by virtue of the artificial gauge field. The asymmetric dispersion shown here is similar to systems where time-reversal symmetry is broken by an actual magnetic field [47,48].

Let us now analyze the spectrum (**Fig. 1c**) of our photonic lattice in detail. For reference, we also plot in **Fig. 1c** the dispersion relation of a two-dimensional array with all the waveguides tilted at the same angle as the core waveguides ($+\eta$) (red-shaded region) and as the cladding waveguides ($-\eta$) (blue-shaded region). In principle, one would expect to have guided modes only in the regions where the dispersion relation of the core does not overlap with that of the cladding, Indeed, we see that this is the case when we plot the spectrum of the photonics lattice (lines in **Fig. 1c**). To better visualize this, we color code the spectrum: the solid blue lines represent propagating modes associated with the entire array (core + cladding), while the red lines represent the two propagating modes confined to the core region. As **Fig 1c** shows, the guided modes reside near the edges of the Brillouin zone $|k_x| \lesssim \pi/d_x$ (but not exactly at $|k_x| = \pi/d_x$), and are nonexistent around the center of the Brillouin zone, $k_x = 0$. As expected, the spectrum is indeed asymmetric with respect to $k_x$, as can be understood by noticing that the symmetric guided mode (top red line in **Fig. 1c**) has a slightly larger support in the $k_x > 0$ region of the spectrum than in the $k_x < 0$ region.

To experimentally demonstrate this guiding phenomenon, we fabricate the photonic lattice sketched in **Fig 1a,b** by using direct laser wiring in fused silica [49,50]. The period in $x$ is $d_x = 16\mu m$, the distance between all the rows along $y$ is $d_y = 22\mu m$, and the tilt angle is $\eta = \pm 8\mu m/cm$. The period is $Z = d_x/\eta = 2cm$ (same parameters used for **Fig 1c**). We fabricate two arrays, one having a core made of two rows, and the other one having a core made of one row only. In both cases, the total propagation distance is $10cm$. In our experiments we launch an elongated Gaussian beam, stretched in $x$ over roughly 10-12 sites, such that it has a well-defined $k_x$-momentum. The specific value of the momentum $k_x$ is controlled by tilting the input beam at the appropriate angle without moving the beam position itself. The input beam has a vertical width corresponding to one row, and is launched into one of the core rows. By measuring the intensity distribution at the output facet, we determine if the input beam is guided inside the core. We do so by calculating the ratio of the power within the core to the total power, measured at the output plane of the lattice, as a function of the launched wavevector $k_x$. The results for the array with two rows in the core are plotted in **Fig 2a**, alongside simulation results using the same parameters (with a wavelength of $\lambda = 633nm$). The experimentally measured profiles of the guiding behavior of the array as a function of the $k_x$ momentum is clearly shown in a movie provided in Supplementary Video 1. We can see that, at $k_x$ values around the edge of the Brillouin zone, almost all of the power remains concentrated in the core – a clear indication that a guided mode exist at those $k_x$ values. In contrast, at $k_x$ values around 0, we see that almost all of the power has escaped from the core into the cladding region. Images of the light at the output facet at specific $k_x$ values are displayed in **Fig 2b** (unguided case) and **Fig 2c** (guided case), where the lack of confinement **(b)** and the waveguiding **(c)** of the light are clearly seen. We obtain similar results using the array with only one row in the core in **Fig. 2 (b-f)** and $\lambda = 532nm$. The experimental measurements of the guiding behavior as a function of the $k_x$ momentum are clearly shown in a movie provided in Supplementary Video 2. Again, clear evidence of guiding at $k_x$ values around the edge of the Brillouin zone is seen both by looking at the ratio of the power at the core to the total power **(Fig. 2d)** for all $k_x$, and by examining the output beam profile at specific $k_x$ values **(Fig. 2e,f)**. Thus, we have experimentally demonstrated, for the first time, light guiding by artificial gauge fields.

Having demonstrated waveguiding by using different artificial gauge fields, we want to take the concept to the next level and ask: would it be possible to guide light in a structure where not only the core and cladding are made from the same material but where they are also subjected to the same gauge field?

To explore this idea, we introduce a different type of gauge field and design waveguides that follow a sinusoidal trajectory. Consider an array of equidistant waveguides, where each waveguide follows a sinusoidal trajectory: $x_{wg}(z) = D\sin(\Omega z + \varphi)$. Here, $\Omega$ is the longitudinal frequency of the sinusoidal modulation, $D$ is the amplitude and $\varphi$ is the phase of the modulation. One can show [29] that the equation of motion for a Bloch wave in this system is

$$i\partial_z \phi_k(z) = 2c_1 \cos\left((k_x + A(z))d_x\right)\phi_k(z), \quad (3)$$

Where $A(z) = k_0 \dot{x}_{wg}(z) = k_0 \Omega D \cos(\Omega z + \varphi)$ is the artificial gauge field associated with the sinusoidal trajectory of the waveguides. Using the high frequency limit ($\Omega \gg c_1$), we can calculate the Floquet spectrum of Eq. (3), and obtain

$$\beta(k_x) = 2c_1 J_0(k_0 \Omega D d_x)\cos(k_x d_x), \quad (4)$$

where $J_0$ is the Bessel function of the first kind. In Eq. (4), we see that the effect of the gauge field is only to renormalize the coupling constant [51]. It is important to note here that the spectrum in

Eq. (4) is completely independent of the phase $\varphi$ of the sinusoidal trajectory of the waveguides. This phase can induce nontrivial effect, for example, in Floquet topological insulators can be used to generate topological defect modes [52,53].

Next, we construct a waveguiding photonic lattice from arrays with sinusoidal trajectories. Consider a two dimensional photonic lattice of such oscillating waveguides, with two rows in the middle acting as the core and the rest acting as cladding. All the waveguides in the array follow sinusoidal trajectories with identical amplitude $D$ and frequency $\Omega$. The cladding waveguides have a trajectory with phase $\varphi = 0$, while the core waveguides have a trajectory with phase $\varphi = \pi$. We calculate the Floquet spectrum of such photonic lattice by Floquet diagonalization of the continuous paraxial equation Eq. (1) [46]. The spectrum is given by **Fig 3c**. It immediately becomes clear that the spectrum of the modes propagating mainly in the core (red lines) is embedded within the spectrum of the cladding (blue lines). This is because, as we mentioned earlier, the spectrum of an array following a sinusoidal trajectory, Eq. (4), does not depend in the phase $\varphi$. Since the dispersion relations of the core and cladding are now exactly the same and therefore fully overlap, we cannot expect to find guided modes such as those presented in **Fig 1b,c** (for the photonic lattice of tilted waveguides). Rather, waveguiding in this sinusoidal structure must rely on a fundamentally new mechanism, different than shifting the dispersion curves of the core with respect to the cladding.

To see the physical origin of guiding in this photonic lattice, we simplify the systems and consider just two arrays with identical gauge fields, but with different oscillation phases. That is, we consider only two adjacent arrays, a top array and a bottom array, where both arrays oscillate along a sinusoidal trajectory with the same amplitude and frequency, but at different phases. Specifically, let the top array have $\varphi = 0$, and the bottom array have $\varphi = \pi$. **See Fig. 3b.** Despite the fact that the two arrays have the same spectrum, the coupling between them is highly nontrivial and z-dependent. To get better understanding, we use the high frequency limit [Supplementary Material], and obtain an approximate expression for the coupling between the two arrays

$$H(k_x) = \begin{pmatrix} 2c_1 J_0(k_0 \Omega D d_x)\cos(k_x d_x) & c_{eff} e^{-\sigma^2 k_x^2/4} J_0(2k_x D) \\ c_{eff} e^{-\sigma^2 k_x^2/4} J_0(2k_x D) & 2c_1 J_0(k_0 \Omega D d_x)\cos(k_x d_x) \end{pmatrix}. \quad (5)$$

According to Eq. (5), one can immediately see that the coupling between the two arrays vanishes at specific $k_x$ values. This means that at values of $k_x$ for which $J_0(2k_x D) = 0$, the top array and bottom array are completely decoupled. At neighboring $k_x$ values the coupling is small, but nonzero.

Now, going back to our full photonic lattice, **Fig. 3a,b**, and examining the modes of the system, we notice that there are indeed modes where the power is mostly concentrated in the core region. When we examine in detail the structure of the modes we find that there are two bands of modes (red bands in **Fig 3c**), which around a certain momentum region (shaded region in **Fig. 3c**) are strongly confined to the core. However, the decay of these confined modes into the cladding region is not exponential, as one would expect from an ordinary guided mode (a bound state in a potential well). Instead, the mode oscillates without any decay, as the mode shown in blue in **Fig. 3d**. This behavior is the hallmark of leaky modes, also known as quantum resonances [54]. However, by recalling the two array system given by Eq. (5) we know that, at specific $k_x$ values, the core and the cladding are completely decoupled in the two array system, and we expect a similar behavior in our full photonic lattice . Indeed, there is a mode at a specific $k_x$ value (red dot in Fig. 3c) that is perfectly confined to the core region, i.e., the penetration of this specific mode into the cladding is identically zero, as shown in red in **Fig. 3d**. If we plot the ratio of peak intensity in the core of these leaky modes to the mean intensity in the cladding, as a function of $k_x$, we find a sharp peak corresponding to $J_0(2k_x R)=0$, where the core and the cladding are completely decoupled (see Supplementary Information). This resonant behavior means that this mode is a Bound State in the Continuum, a BIC. Conventionally, BICs are bound modes whose energy is embedded in the continuum part of the spectrum, but they are nevertheless forbidden from coupling to the continuum modes due to special symmetries or the topology they possess [40–44]. Here, we show that our guided mode corresponds to a dynamic BIC [42], which has thus far never demonstrated in experiment.

To experimentally demonstrate this new guiding phenomenon, we fabricate the photonic lattice of **Fig 3a** by using direct laser wiring in fused silica [49,50]. The period in $x$ is $d_x = 20 \mu m$, the distance between the arrays along $y$ is $d_y = 24 \mu m$, the frequency is $\Omega = 2\pi \, rad/cm$, and the amplitude is $D = 8 \mu m$ (same parameters used in **Fig 3c**). We launch an elongated Gaussian beam with well-defined $k_x$ into the core, and measure the ratio of the power within the core to the total power measured at the output plane of the lattice. The results are plotted in **Fig. 4a**, which shows two peaks at $k_x = 0.73\pi/d_x$, clearly demonstrating that the core of our photonic lattice is guiding light at a specific value of momentum. Notice that the expected peak positions from continuum simulations (arrows in **Fig. 4a**) are in perfect agreement with our measurements. The resonant nature of the guiding mechanism can clearly be observed at the measured beam profile at the output facet for a $k_x$ corresponding to maximum guiding, **Fig. 4b**, and for a $k_x$ where there is no guiding, **Fig. 4c**. The experimental measurements of the guiding behavior as a function of the $k_x$ momentum over the entire Brillouin zone are clearly shown in a movie provided in Supplementary Video 3.

To fully characterize this phenomenon, we design a set of experiments to measure how the position of maximum guiding (the BIC's position) is affected by changing the amplitude $D$ of the sinusoidal trajectory. According to our approximation in Eq. (5), the peak position should be decreasing with increasing $D$. To corroborate this, we fabricate three photonic lattices – with sinusoidal amplitudes $D = 6\mu m, 8\mu m,$ and $10\mu m$. We again launch an elongated Gaussian beam with well-defined $k_x$ into the core and measure the ratio of power at the core; the results are shown in **Fig 5**. We clearly see that the position of the peak where the guiding occurs (i.e. the BIC position) decreases for lattices with increasing sinusoidal amplitude $D$, as we predicted from our theoretical analysis. Our experimental results also match very well the full-wave continuum simulations, as marked by the arrows in **Fig. 5**. Altogether, the experimental result (Figs. 3-5) demonstrate a new gauge-induced waveguiding mechanism, where the core and the cladding regions are made from the same material, and are subjected to the same artificial gauge fields characterized by the same dispersion curves, with the only difference being a phase shift in the gauge field. Waveguiding of this sort relates to BICs, quantum resonances and dynamic localization, and offers a new platform for engineering these phenomena by making use of artificial gauge fields.

In summary, we experimentally demonstrated the phenomenon of guiding light by virtue of artificial gauge fields. We fabricate photonic lattices where artificial gauge fields induce different dispersion relations at the core and cladding regions, and observed how this gives rise to guided modes. In a more advanced setting, we demonstrated how to induce waveguiding by only changing the phase of the gauge field, while the dispersion curves of the core and clad are identical. We showed that in such settings, the guiding occurs because the interaction between the core and the cladding is nontrivial and can lead to nulling the coupling altogether, which offers precise control over the momentum values at which the guiding takes place. The new guiding mechanisms demonstrated here open the door to new applications of artificial gauge fields in photonics, and by their fundamental nature are applicable not only to the entire electromagnetic spectrum and different optical systems, but also to other physical systems such as acoustics and cold atoms.

**Acknowledgements:** this work was supported by the German-Israeli DIP Program (grant BL 574/13-1), by the United States Air Force Office of Scientific Research, and by Deutsche Forschungsgemeinschaft (grants SZ 276/9-1, SZ 276/19-1, SZ 276/20-1). The authors would like to thank C. Otto for preparing the high-quality fused silica samples used in all experiments presented here.

**Authors Contributions:** all authors contributed significantly to this work.

## Methods

**Fabrication.** We fabricated the waveguides in 10cm fused silica glass (Corning 7980) samples by the femtosecond laser writing method [49]. We used pulses created by a Coherent RegA seeded with a Mira 900 have an energy of 450nJ at 800nm and 100kHz. An Aerotech ALS 130 together with a microscope objective (0.35NA) provides the highly accurate focusing of the laser beam 50 μm to 800 μm under the sample surface. By moving the sample with a speed of 100 mm/min the refractive index at the focal point is changed around $7 \cdot 10^{-4}$. This creates the waveguides with a mode field diameter of 10.4 μm x 8 μm at 632.8 nm. Propagation losses and birefringence were estimated at 0.2dB cm$^{-1}$ and $10^{-7}$, respectively. A table with all the parameters and microscope images of the fabricated optical lattices is given in the supplementary material.

**Experimental Setup.** We generate an elongated Gaussian beam ($\sim 130 - 190 um$ $x$-width and $\sim 8 - 12 um$ $y$-width at $\lambda = 633 nm$ and $532 nm$) at the input facet of the waveguide array by taking the Fourier transform of a rectangular slit using a single achromatic lens. The position of the slit (Fourier plane of the elongated Gaussian beam) is scanned in the $x$-direction ($\sim 10 um$ per step) by using a stepper motor to scan the phase of the input beam, which is translated into scanning the $k_x$ of the Bloch mode launched into the photonic lattice. This allows us to collect up to ~800 measurements inside the Brillouin zone. The light intensity distribution at the output facet of the array is image onto a linear CMOS camera (1920 × 1200 pixels, 1/1.2" sensor) by using $10 \times$ and $20 \times$ objectives. The size of the Brillouin Zone (or the $k_x$ momentum step) is calculated using geometric optics and by correlating the output profiles.

**Simulations.** All the simulation results are obtained from full continuum simulations of the paraxial wave equation with the corresponding lattice refractive index profile. Each waveguide is modeled as a super-Gaussian profile given by $\Delta n = \Delta n_0 exp\left(-(x/w_x)^6 - (y/w_y)^6\right)$ with $w_x = 1.9 \mu m$, $w_y = 5.5 \mu m$ and $\Delta n_0 = 7.5 \cdot 10^{-4}$. The background refractive index is $n_0 = 1.45$. These parameters were estimated from the output profile of single, 1D and 2D arrays of straight waveguides that were fabricated together with the lattices study here. The continuum Floquet band structure and eigenmodes presented in Fig 1c, Fig. 3c,d are calculated using a continuum Floquet eigensolver described in the supplementary material of Ref. [46].

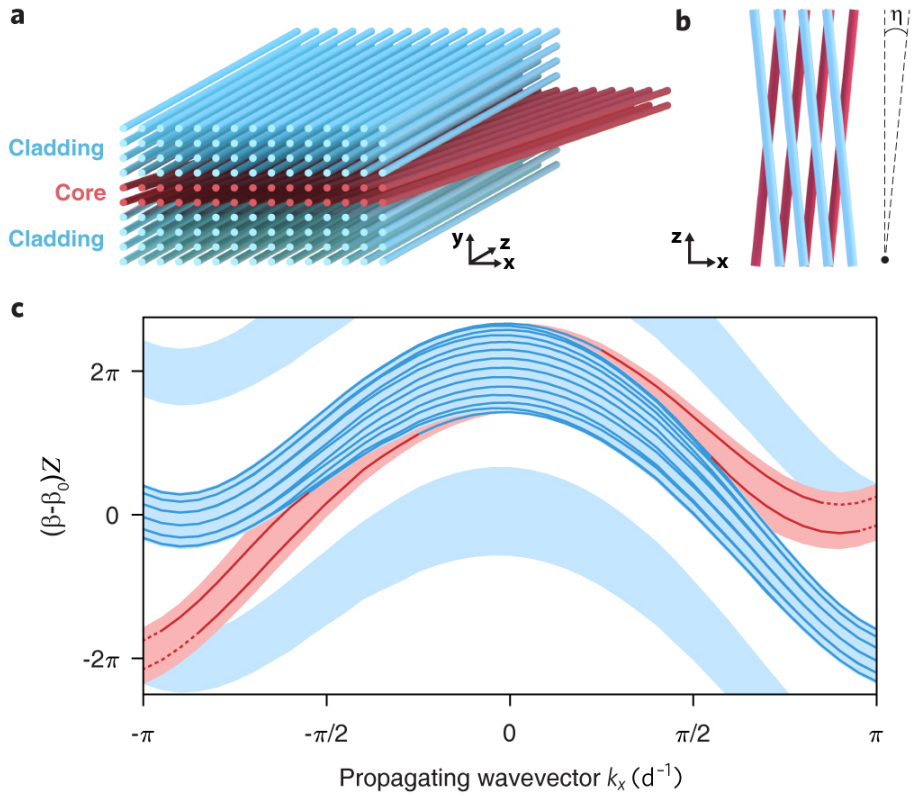

**Figure 1 | Waveguiding of light by artificial gauge fields in tilted arrays. (a), (b)** Schematic of the two-dimensional waveguide array displaying waveguiding by artificial gauge field. The cladding rows (blue) are tilted to the left (at an angle $-\eta$) and the two rows at the center (red) are tilted to the right (at an angle $+\eta$), thereby serving as the guiding core. The tilt angle introduces an artificial gauge field, which is different for the core and cladding regions. **(c)** Dispersion relation of the waveguide array depicted in (a). The z-periodicity of the structure gives rise to a periodic dispersion relation (described by a Floquet spectrum). The blue-shaded sections represent the dispersion relation of the cladding modes (the arrays tilted by $-\eta$) while the red-shaded section represents the dispersion relation of just the two core arrays tilted at $+\eta$. The solid blue lines represent propagating modes associated with the whole array (core + cladding) while the red lines represent the two propagating modes confined to the core region. Around $|k_x| = \pi/d_x$, the first Floquet bands of the core intercept the second Floquet bands of the cladding, creating a weak coupling between both regions (dashed red lines). The distance between the waveguides is $d_x = 16\mu m$ and $d_y = 22\mu m$, the tilting angle is $\eta = \pm 8\mu m/cm$. The spectrum is calculated using $\lambda = 633\mu m$.

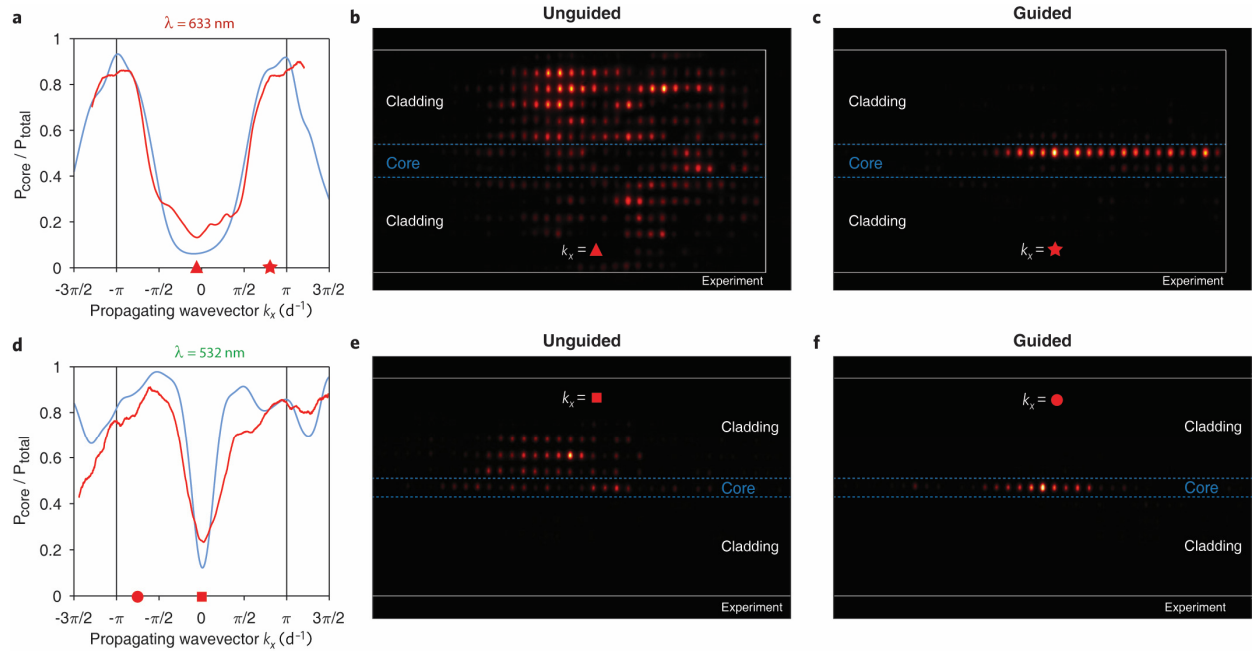

**Figure 2 | Experimental results displaying waveguiding by artificial gauge fields in tilted arrays.** We launch, into one of the core arrays, an elongated Gaussian beam tilted in the x-direction such that we control its wavenumber $k_x$. **(a)** Ratio of the power at the core to the total power measured at the output plane of the lattice, as a function of the launch wavevector, $k_x$, as calculated in continuum simulation (blue) and as measured in the experiment (red). **(b),(c)** Experimentally measured intensity pattern observed at the lattice output for tilt wavevector (b) in the unguided region ($k_x$ marked in (a) with a tringle), and (c) in the guiding region ($k_x$ marked in (a) as a start). **(d-f)** Same as (a-c) for but for a single-core tilted lattice.

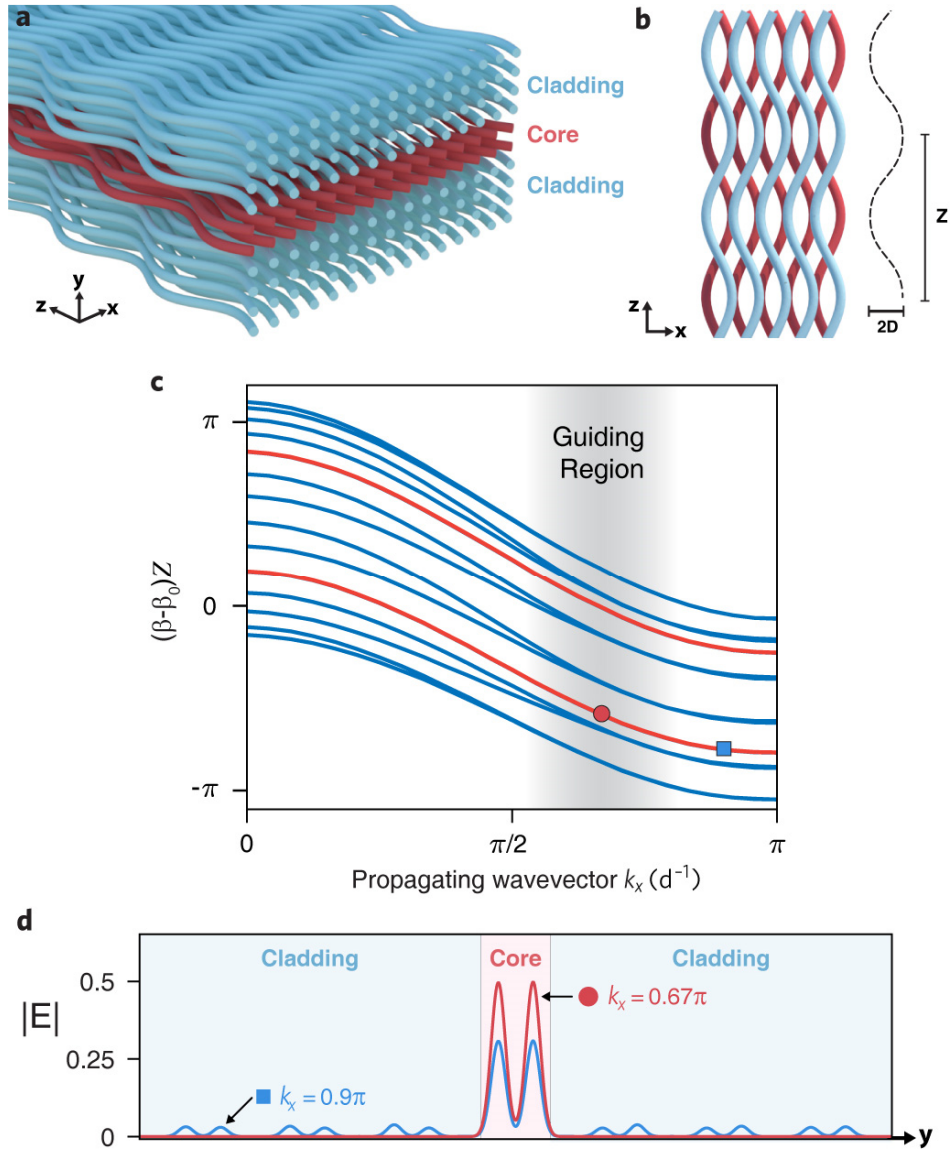

**Figure 3 | Waveguiding of light by phase-shifted but otherwise identical artificial gauge fields. (a),(b)** Schematic of the two dimensional waveguide array of sinusoidal trajectories displaying guiding by phase-shifted artificial gauge fields. All the waveguides have the same index profile and follow a sinusoidal trajectory with the same periodicity and amplitude (A), but the cladding (blue) and core (red) rows are shifted by a $\pi$ phase. **(c)** Dispersion relation of the waveguide array depicted in (a). The z-periodicity of the structure gives rise to periodic dispersion relation (described by a Floquet spectrum). The blue lines represent the modes that are extended over the whole array, while the red lines represent the modes confined to the core waveguides.

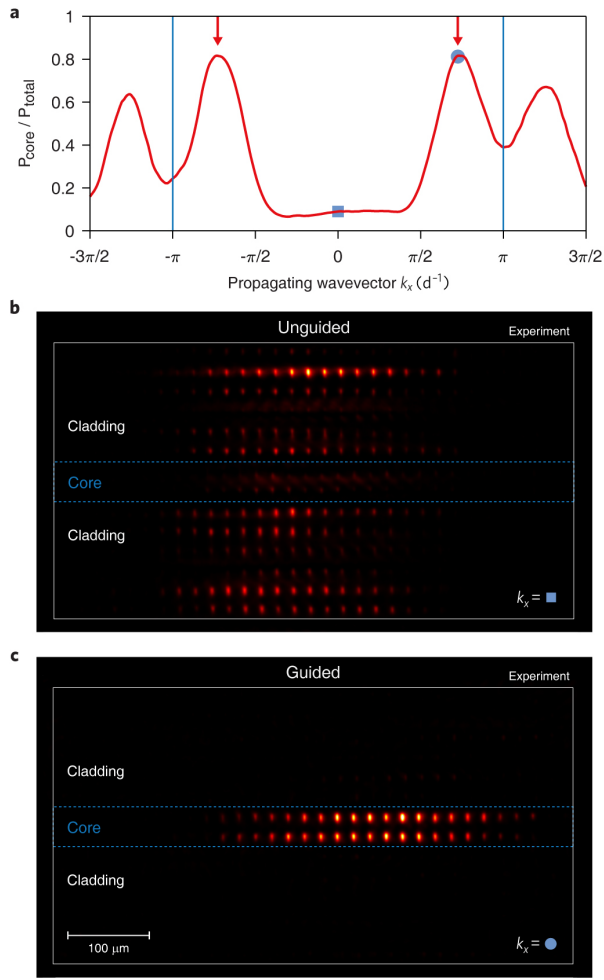

**Figure 4 | Experimental results displaying waveguiding by phase-shifted artificial gauge fields.** We launch, into both of the core arrays, an elongated Gaussian beam tilted in the x-direction such that we control its wavenumber $k_x$. The lattice oscillation amplitude is $8\mu m$. **(a)** Ratio of the power at the core to the total power in the array, measured at the output plane of the lattice, as a function of the launch wavevector, $k_x$. The red arrows mark the position of maximum core guiding calculated by continuum simulation, which clearly agrees with the experimental results. The blue horizontal lines mark the Brillouin zone (notice the expected peak replicas outside the Brillouin zone). **(b)**, **(c)** Experimentally observed intensity profile at the lattice output for a propagating wavevector in (a) the unguided region ($k_x$ marked in (a) with a square) and in the (c) guiding region ($k_x$ marked in(a) as a circle).

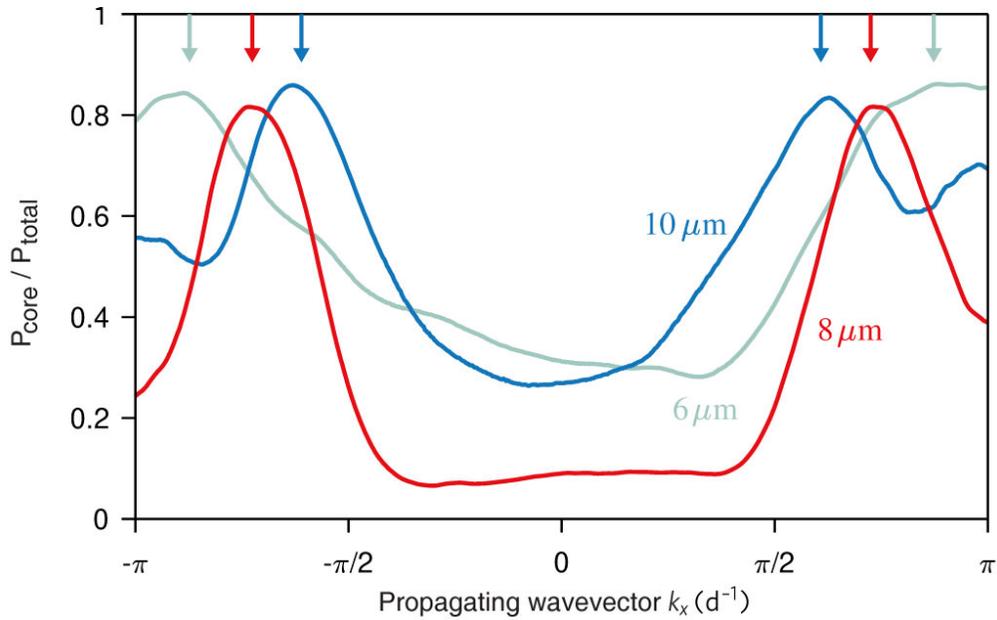

**Figure 5 | Experimental results displaying phase-shifted gauge guiding in sinusoidal arrays.** We launch an elongated Gaussian beam ($\lambda = 633nm$) into one of the core rows and scan its tilt angles in the x-direction, thereby scanning the $k_x$ momentum component. The figure shows the experimentally measured ratio of the power in the core to the total power in the array, measured at the output plane of the lattice, as a function of the launched wavevector, $k_x$, for three different lattices with different oscillation amplitudes: $6\mu m$ (green), $8\mu m$ (red), and $10\mu m$ (blue). Notice how the peak position of the guided power moves to smaller $|k_x|$ as we increased the oscillation amplitude of the array. The arrows mark the position of maximum core guiding calculated by continuum simulation, which clearly agree with the experimental results.